\documentclass[aps,prl,twocolumn,showpacs,groupedaddress]{revtex4-1}  
\usepackage{graphicx}  
\usepackage{graphics}
\usepackage{dcolumn}   
\usepackage{bm}        
\usepackage{amssymb}   

\begin{document}



\title{Intrinsic noise in stochastic models of gene expression 
with molecular memory and bursting}
\author{Tao Jia}
\email{tjia@vt.edu}
\author{Rahul V. Kulkarni}
\email{kulkarni@vt.edu} 
\affiliation{Department of Physics, \\
Virginia Polytechnic Institute and State University, \\
Blacksburg, VA 24061}
\date{\today}

\begin{abstract}
Regulation of intrinsic noise in gene expression is essential for many
cellular functions. Correspondingly, there is considerable interest in
understanding how different molecular mechanisms of gene expression
impact variations in protein levels across a population of
cells. In this work, we analyze a stochastic model of bursty gene
expression which considers general waiting-time distributions governing
arrival and decay of proteins. By mapping the system to models
analyzed in queueing theory, we derive analytical expressions
for the noise in steady-state protein distributions. The derived
results extend previous work by including the effects of arbitrary
probability distributions representing the effects of molecular memory
and bursting. The analytical expressions obtained provide insight into
the role of transcriptional, post-transcriptional and
post-translational mechanisms in controlling the noise in gene
expression.
\end{abstract}

\maketitle 


Regulation of gene expression is at the core of cellular adaptation
and response to changing environments. Given that the underlying
processes are intrinsically stochastic, cellular regulation must be
designed to control variability (noise) in gene expression
\cite{kaern05}. While noise reduction is essential in many cases,
regulatory mechanisms can also exploit the intrinsic stochasticity to
increase noise and generate phenotypic heterogeneity in a clonal
population of cells \cite{raj08}. Quantifying the contributions of
different sources of intrinsic noise using stochastic models of gene
expression \cite{paulsson05,azaele09,munsky09} is thus an important
step towards understanding cellular processes and variations in cell
populations.

Several recent studies have focused on quantifying noise in gene
expression. Experiments have shown that protein production often
occurs in `bursts' \cite{cai06,yu06} 
and single-molecule measurements have also provided
evidence for transcriptional bursting, i.e.\ production of mRNAs in
bursts \cite{golding05,raj06,chubb06}. The 
analysis and interpretation of such experimental studies
has been aided by the development of coarse-grained
stochastic models of gene expression. The simplest of these considers
the basic processes (transcription, translation and degradation) as
elementary Poisson processes \cite{thattai01} with exponential
waiting-time distributions. However, since these processes are known
to involve multiple biochemical steps, the corresponding waiting-time
distributions can be more general than the `memoryless' exponential
distribution\cite{pedraza08}.  An important question then arises: how do gene
expression mechanisms involving molecular memory effects influence the
noise in protein distributions?

\begin{figure}[tb]
\begin{center}
\resizebox{6.5cm}{!}{\includegraphics{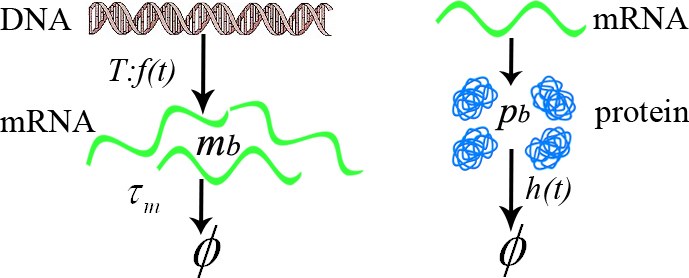}}
\caption{Reaction scheme for the underlying gene expression model.
Production of mRNAs occurs in bursts (characterized by
random variable $m_b$ with arbitrary distribution) and each mRNA gives
rise to a burst of proteins (characterized by random variable $p_b$
with arbitrary distribution) before it decays (with lifetime $\tau_m$). 
The waiting-time distributions for burst arrival and decay of proteins are
characterized by the functions $f(t)$ and $h(t)$ respectively. 
\label{fig1}}
\end{center}
\end{figure}\noindent

Motivated by the preceding observations, we introduce a model including
general waiting-time distributions for processes governing the arrival
of bursts and the decay of proteins (termed `gestation' and
`senescence' effects respectively \cite{pedraza08}). 
The underlying reaction scheme for the models analyzed in this work is
shown in Fig. 1. Production of mRNAs occurs in independent
bursts and the time interval between the
arrival of consecutive mRNA bursts is characterized by random variable $T$
with corresponding probability density function (p.d.f) $f(t)$.
The number of mRNAs produced in a single transcriptional burst is
characterized by the random variable $m_b$.  Each mRNA independently
gives rise to a random number of proteins (characterized by random
variable $p_b$) before it is degraded. For the basic models of translation, 
$p_b$ follows the geometric distribution \cite{cai06,yu06,friedman06}. 
However, more general schemes of gene expression 
(e.g.\ involving post-transcriptional
regulation \cite{jia10}) can give rise to protein burst distributions
that deviate significantly from a geometric distribution. 
Proteins are degraded independently and the waiting-time distribution
for protein decay is characterized by the p.d.f $h(t)$.

In the limit that the mRNA lifetime ($\tau_m$) is much shorter than
the protein lifetime ($\tau_p$), i.e.\ $\frac{\tau_m}{\tau_p} \ll 1$, 
the evolution of cellular protein
concentrations can be modeled by processes governing arrival and decay
of proteins alone \cite{friedman06,swain08}.  Unless otherwise
stated, the analysis in this paper will focus on this `burst' limit,
in which proteins are considered to arrive in independent
instantaneous bursts arising from the underlying mRNA burst.  In this
limit, we have shown in recent work \cite{elgart09} that the
processes involved in gene expression can be mapped on to models analyzed
in queueing theory. In this mapping, individual proteins are the
analogs of customers in queueing models. The bursty synthesis of
proteins then corresponds to the arrival of customers in `batches',
whereas the protein decay-time distribution is the analog of the
service-time distribution for each customer. Given that degradation of
each protein is independent of others in the system, the process maps
on to queueing systems with infinite servers.
Correspondingly, the gene expression model in Fig. 1 maps on
to what is known as a $GI^{X}/G/\infty$ system in the queueing
literature.  In this notation, the symbol $G$ refers to the general
waiting-time distribution and $I^{X}$ indicates that the customers
arrive in batches of random size $X$, where $X$ is drawn independently
each time from an arbitrary distribution.

The $GI^{X}/G/\infty$ system has been analyzed in previous work in
queueing theory \cite{liu00}. In the following, we briefly review the
notation and relevant results from the queueing theory analysis. As in
Fig. 1, $f(t)$ and $h(t)$ denote the p.d.f. for the arrival time and
service time respectively, with $F(t)$ and $H(t)$ as the corresponding
cumulative density functions (c.d.f).  The distribution of batch size
$X$ has the corresponding generating function $A(z)$, defined as
$A(z)=\sum_{i=1}^{\infty}{P(X=i) z^i}$.  The $k$th factorial moment of
batch size $X$, denoted by $A_k$, is given by
$A_k=(d^kA(z)/dz^k)|_{z=1}$.  The number of customers in service at
time $t$ is denoted by $N(t)$ and analytical expressions have been
derived for the $r^{\mathrm{th}}$ binomial moment $B_r(t)$ of $N(t)$
\cite{liu00}. These results can be used to derive expressions for all
the moments of $N(t)$, for example $E[N(t)] = B_1(t)$ and $Var[N(t)] =
2B_2(t) + B_1(t) -B_1^2(t)$. In the following, we will focus on two
general subcategories of the $GI^{X}/G/\infty$ system for which
closed-form analytical expressions can be derived for the mean and
variance of steady-state protein distributions. These correspond to
two cases: A) arbitrary distributions for gestation and bursting
with a Poisson process governing protein degradation and B) arbitrary
distributions for bursting and senescence with a Poisson process
governing burst arrival.

Consider first case A, for which arbitrary gestation and bursting
effects are included. In this case, the random variable $T$
characterizing the time interval between bursts is drawn from an
arbitrary p.d.f. $f(t)$.  The protein decay-time distribution $h(t)$
is taken to be an exponential function with $h(t)=\mu_p e^{-\mu_p t}$
and the mean protein lifetime is given by $\tau_p = 1/\mu_p$.  The
corresponding queueing system is $GI^X/M/\infty$ where $M$ indicates
that the process of customer departure, which is the analog of protein
decay, is Markovian.  $A(z)$ corresponds to the generating function of
burst size distribution (determined by random variables $m_b$ and
$p_b$ in Fig. 1) and $N(t)$ denotes the number of proteins in the cell
at time $t$.  The previous analysis \cite{liu00} has derived
expressions for the steady-state mean and variance corresponding to
$N=\lim_{t \to \infty}N(t)$ for the $GI^X/M/\infty$ queue as
\cite{error_note}:
\begin{eqnarray}
E[N] &=& \frac{1}{\mu_p \langle T\rangle}A_1  \nonumber \\
Var[N] &=& E[N] (1 + \frac{f_L(\mu_p)}{1-f_L(\mu_p)} A_1 - E[N] + \frac{A_2}{2A_1}), \label{eq:queue_2} 
\end{eqnarray}
where $\langle T\rangle$ is the mean of p.d.f $f(t)$ and $f_L(s)$ is
the Laplace transform of $f(t)$.

To translate the result Eq.(\ref{eq:queue_2}) into an expression for
the noise in protein distributions, we derive expressions for $A_1$
and $A_2$ in terms of variables characterizing mRNA and protein burst
distributions.  In general, each mRNA will produce a random number of
proteins ($p_b$) and furthermore the number of mRNAs in the burst is
also a random variable ($m_b$). The number of proteins produced in a
single burst is thus a compound random variable.
Correspondingly, using standard results from probability theory \cite{ross}, we
derive the following equations for burst size parameters ($A_1$ and
$A_2$) in terms of $m_b$ and $p_b$:
\begin{eqnarray}
A_1&=&\langle m_b\rangle \langle p_b\rangle  \nonumber \\
A_2&=&\langle m_b\rangle  (\sigma^2_{p_b} - \langle p_b\rangle) + (\sigma_{m_b}^2 + \langle m_b\rangle ^2) \langle p_b\rangle ^2  \label{eq:A2},
\end{eqnarray}
where the symbols $\langle..\rangle$ and $\sigma$ represent the mean and 
standard deviation respectively.

Using Eq.(\ref{eq:A2}), in combination with identification of the
random variable $N$ with the corresponding variable characterizing the
number of proteins ($p_s$), we obtain the following
expressions for the mean and coefficient of variance (noise) of the
steady-state protein distribution:
\begin{eqnarray}
\langle p_s\rangle &=&  \frac{\tau_p}{\langle T\rangle} \langle m_b\rangle \langle p_b\rangle \nonumber \\
\frac{\sigma^2_{p_s}}{\langle p_s\rangle ^2} &=& \frac{1}{\langle p_s\rangle } + \frac{\langle T\rangle}{2 \tau_p} \times \Bigl( K_g + \sigma^2_{m_b}/\langle m_b\rangle ^2 \nonumber \\
&+& \frac{\sigma^2_{p_b}/\langle p_b\rangle ^2 - 1/\langle p_b\rangle}{\langle m_b\rangle} \Bigr), \label{eq:exact}
\end{eqnarray}
where 
\begin{equation}
K_g = 2\Bigl( \frac{f_L(\mu_p)}{1-f_L(\mu_p)} - \frac{1}{\mu_p\langle T\rangle} \Bigr) +1 \label{eq:K}, 
\end{equation}
is denoted as the {\it gestation factor}.

\begin{figure}[tb]
\begin{center}
\resizebox{8.5cm}{!}{\includegraphics{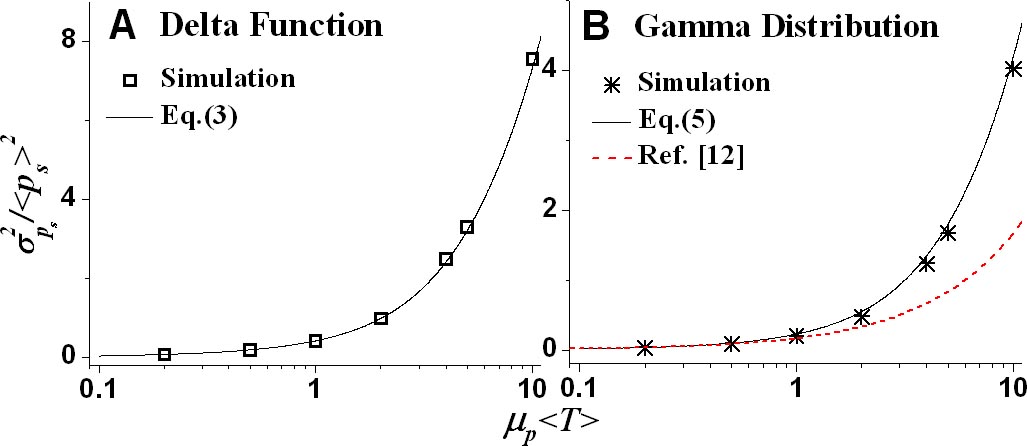}}
\caption{The noise {\it vs} $\mu_p\langle T\rangle$ from analytical
expressions and stochastic simulations. A) The time interval between consecutive bursts
is fixed and only 1 mRNA is produced each burst.
The protein production is under post-transcriptional 
regulation \cite{jia10} such
that $\sigma^2_{p_b} = 0.67 \langle p_b\rangle ^2 +\langle
p_b\rangle$ and $\tau_m/\tau_p \approx 0.02$.  B) The time
interval between bursts is drawn from a Gamma distribution and the
number of mRNAs created in one burst is drawn from a Poisson
distribution. The number of proteins created by each mRNA follows a
geometric distribution. The parameters are $\tau_m/\tau_p = 0.2$,
$\langle m_b\rangle=10$, $\sigma^2_{m_b}/\langle m_b\rangle^2=0.1$ and
$\sigma^2_{T}/\langle T\rangle^2=0.2$.  While
Eq.(\ref{eq:approximate}) agrees with simulations, the
result from Ref. \cite{pedraza08} is less accurate when $\mu_p\langle T\rangle$
is large.
\label{fig2}}
\end{center}
\end{figure}\noindent

Different contributions to the noise in protein distributions are
highlighted in Eq.(\ref{eq:exact}): gestation effects, mRNA
transcriptional bursting, and translational bursting from a single
mRNA, which correspond to the terms $K_g$, $\sigma^2_{m_b}/\langle
m_b\rangle ^2$ and $\sigma^2_{p_b}/\langle p_b\rangle ^2$,
respectively. The first two terms can be modified by transcriptional
regulation and the last term can be tuned by post-transcriptional
regulation. It is noteworthy that each source contributes additively
to the overall noise in the steady-state distribution. Moreover, while
the noise due to gestation effects is independent of the degree of
transcriptional bursting, the noise contribution from translational
bursting is effectively reduced by transcriptional bursting.

While Eq.(\ref{eq:exact}) is valid for general gestation effects, it
is of interest to consider specific examples.  We consider the case
such that there is a constant delay between arrival of consecutive mRNA
bursts, i.e.\ the waiting-time distribution is
$f(t) = \delta(t- T_d)$. In this case, the gestation factor
is given by $K_g = 2e^{-\mu_p T_d}/(1-e^{-\mu_p T_d}) - 2/\mu_p T_d +
1$. The corresponding expression for the noise in protein
distributions Eq.(\ref{eq:exact}), considering a general case which
also includes the effects of post-transcriptional regulation
\cite{jia10}, is in excellent agreement with results from stochastic
simulations (Fig. 2A).  It is noteworthy that $K_g$ can be
nonvanishing even though the time interval between consecutive bursts
is fixed (i.e.\ $\sigma^2_{T} =0$). In contrast to previous work
\cite{pedraza08}, which suggests that the contribution of gestation
effects to the noise vanishes when $\sigma^2_{T} =0$, our result shows
that $K_g$ can be tuned from 0 to 1 as $\mu_p T_d$ is
varied.

While the results derived above are valid in the limit $\tau_m \ll \tau_p$,
an exact expression for the noise in the general case (i.e.\ without invoking 
the condition $\tau_m \ll \tau_p$ and for general gestation and bursting
distributions) is difficult to obtain. 
However, a useful approximation can be obtained by noting that, for 
the basic gene expression models, the exact result
is obtained by scaling the terms in the bracket in Eq.(\ref{eq:exact}) 
with a time-averaging factor
$\frac{\tau_p}{\tau_m+\tau_p}$ \cite{paulsson05, bareven06}.
Using the approximation that the time-averaging
factor is the same for general gestation and bursting distributions, we obtain
\begin{eqnarray}
\frac{\sigma^2_{p_s}}{\langle p_s\rangle ^2} &\approx& \frac{1}{\langle p_s\rangle } + \frac{\langle T\rangle}{2 \tau_p} \times \Bigl( K_g + \sigma^2_{m_b}/\langle m_b\rangle ^2 \nonumber \\
&+& \frac{\sigma^2_{p_b}/\langle p_b\rangle ^2 - 1/\langle p_b\rangle}{\langle m_b\rangle} \Bigr)\times\frac{\tau_p}{\tau_m+\tau_p}, \label{eq:approximate}
\end{eqnarray}

It is instructive to compare Eq.(\ref{eq:approximate}) with the
result derived in previous work \cite{pedraza08} 
which assumes the basic protein production reaction scheme
such that $\sigma^2_{p_b} = \langle p_b\rangle ^2 +\langle p_b\rangle$.
Considering this specific case,
we note that Eq.(\ref{eq:approximate}) is identical to the previous result 
\cite{pedraza08} apart from the terms corresponding
to the gestation factor $K_g$.
The connection to the previous result can be seen by expanding the Laplace transform, $f_L(\mu_p)$, in terms of moments of $T$.
By assuming $\mu_p\langle T\rangle$ is small and $\langle T^n\rangle$  
scales as the $n^{th}$ power of $\langle T\rangle$ or less,
$K_g$ can be approximated by 
$K_g \approx \sigma^2_T/\langle T\rangle ^2$ 
which corresponds to the previous result. Since the parameter 
$1/(\mu_p\langle T\rangle)$ measures the mean number of bursts occurring 
during the  protein lifetime, this indicates that the previous result
\cite{pedraza08} is valid for
the case of frequent bursting during a protein lifetime, 
and breaks down when bursts occur over larger time intervals (Fig. 2B).

We now consider case B, which corresponds to arbitrary distributions
for bursting and senescence effects along with exponential
waiting-time distributions for burst arrival. 
For this case, 
we take the waiting-time for protein degradation to be drawn from an
arbitrary distribution characterized by p.d.f $h(t)$ and c.d.f
$H(t)$. The waiting-time between consecutive bursts is characterized by  
an exponential distribution with $f(t) = \lambda e^{-\lambda t}$. 
The corresponding system, following the mapping to queueing
theory, is the $M^{X}/G/\infty$ queue. The steady-state mean and
variance of $N$ for this queue has been obtained in previous work \cite{liu00}:
\begin{eqnarray}
E[N]&=&\lambda A_1 \int_{0}^{\infty}[1-H(t)]dt \nonumber \\
Var[N]&=&E[N] + \lambda A_2 \int_{0}^{\infty}[1-H(t)]^2dt \label{eq:queue_3}.
\end{eqnarray}

By taking Eq.(\ref{eq:A2}) and the relation $\langle T\rangle = 1/\lambda$ 
into account, the mean and the noise for arbitrary senescence and bursting 
distribution can be derived as:
\begin{eqnarray}
\langle p_s\rangle &=& \frac{A_1}{\langle T\rangle} \int_{0}^{\infty}[1-H(t)]dt = \frac{\tau_p}{\langle T\rangle} \langle m_b\rangle \langle p_b\rangle \nonumber \\
\frac{\sigma^2_{p_s}}{\langle p_s\rangle ^2} &=& \frac{1}{\langle p_s\rangle } + \frac{\langle T\rangle}{2\tau_p} \times \Bigl( 1 + \sigma^2_{m_b}/\langle m_b\rangle^2 \nonumber \\
&+& \frac{\sigma^2_{p_b}/\langle p_b\rangle ^2 - 1/\langle p_b\rangle}{\langle m_b\rangle} \Bigr) \times K_s, \label{eq:senescence}
\end{eqnarray}
where 
\begin{equation}
K_s = \frac{2 \int_{0}^{\infty}[1-H(t)]^2dt}{\tau_p} = 2- \frac{2 \int_{0}^{\infty}H(t)[1-H(t)]dt}{\tau_p},\label{eq:Ks}
\end{equation}
is denoted as the {\it senescence factor}.

It is noteworthy Eq.(\ref{eq:senescence}) and Eq.(\ref{eq:exact}) have
multiple terms in common. The terms characterizing the noise from
transcriptional and translational bursting remain unchanged. 
However, unlike the gestation factor that contributes to
the total noise {\it additively}, the senescence factor serves as a
{\em scaling} factor for the total noise. While there is no obvious
upper limit on the value of $K_g$, the upper bound for $K_s$ is 2 as
is evident from Eq.(\ref{eq:Ks}). In general, as the distribution
$h(t)$ grows more sharply peaked, the $K_s$ value increases.  When
$h(t)$ becomes a delta function, $K_s$ reaches its maximum value.

The general results derived in this work will serve as useful 
inputs for the analysis and interpretation of diverse experimental studies 
of gene expression. Some examples are: 1)
Recent experiments on single-cell studies of HIV-1 viral infections
have focused on the frequency and degree of transcriptional bursting 
\cite{Skupsky10}.
For such studies, the derived results can be used to 
relate measurements of inter-arrival
waiting-time distributions and burst distributions to the noise in
protein distributions. 
2) Experimental data and  computational models of the cell-cycle in
 yeast indicate that modeling the basic processes of gene expression 
as Poisson
processes gives rise to unrealistically large noise in protein 
distributions \cite{tyson09},
thereby suggesting that regulatory schemes which change distributions to
reduce the noise are employed by the cell. The analytical
expressions derived highlight different contributions to noise and can thus
provide insight into how different regulatory schemes can lead to noise
reduction. 
3) More generally, the results derived can be used in the analysis of 
inverse problems, i.e. using experimental measurements of intrinsic 
noise to determine parameters of the underlying 
kinetic models. Such efforts, in turn, can lead to further insights into 
cellular factors that impact gene regulation, based on experimnetal 
observations of noise in gene expression.

In summary, we have analyzed the noise in protein distributions for general
stochastic models of gene expression. The present
work extends previous analysis by deriving analytical results for the noise 
in protein distributions for arbitrary gestation, senescence and bursting 
mechanisms. The expressions obtained provide insight into 
how different sources contribute to the noise in protein levels which can lead 
to phenotypic heterogeneity in isogenic populations. The results derived will
thus serve as useful inputs for the analysis and interpretation of 
experiments  probing stochastic gene expression and its 
phenotypic consequences. 
At a broader level, this work demonstrates the 
benefits of developing a mapping between models of stochastic gene expression
and queueing systems which has potential applications for research in both 
fields. The extensive analytical approaches and tools developed in queueing 
theory can now be employed to analyze stochastic processes in gene expression.
It is also anticipated that future analysis of regulatory mechanisms 
for gene expression will lead to new problems and challenges for queueing
theory.


The authors acknowledge funding support from  NSF (PHY-0957430) and from ICTAS, Virginia Tech. 

\bibliographystyle{apsrev}
\bibliography{stochastic_modeling}

\begin{thebibliography}{22}
\bibitem[{err()}]{error_note} The result given in Ref. [17] has a minor error which is corrected here.
\end{thebibliography}

\end{document}